\documentclass[twocolumn, prl]{revtex4}
\pdfoutput=1
\usepackage{graphicx}
\usepackage{epsfig}
\begin{document}
\title{Universal Cooling Scheme for Quantum Simulation} 
\author{Tin-Lun Ho$^{\dagger}$ and Qi Zhou$^{\dagger\dagger}$}
\affiliation{$^{\dagger}$Department of Physics, The Ohio State University, Columbus, OH 43210
\\$^{\dagger\dagger}$Joint Quantum Institute and Condensed Matter Theory Center, Department
of Physics, University of Maryland, College Park, MD 20742}
\date{\today}
\begin{abstract} 
At present, there is a worldwide effort to use cold atoms to simulate strongly correlated quantum many-body systems. It is hoped that these ``simulations" will provide solutions to many unsolved problems.  However, the relevant energy scales in most of these experiments are so small that one has to go to  entropy regimes far below those achievable today.  Here, we present a general scheme to extract entropy directly from the region of interest. The late stage of this process is equivalent to a continuous ``evaporation", and is able to combat intrinsic heating of the system.  
For illustration, we show how to cool a weak coupling BCS superfluid (with  $T_{c}\sim 10^{-10}$K) to $10^{-11}$K with this simple procedure, with entropy per particle as low as $5\times 10^{-4}k_{B}$ in the superfluid region.

 \ \end{abstract}
\maketitle

Although quantum simulations using cold atoms are considered as means to find solutions to unsolved many-body problems, they are in fact an exceedingly ambitious undertaking in ultra-low temperature physics. For example, the antiferromagnetic phase of the 3D fermion Hubbard model is characterized by the virtual hopping scale $t^2/U$, which can be $10^{-11}$K or lower\cite{HoZhouheating}. 
In terms of entropy per particle, it is estimated that one has to lower the current experimentally achievable value by about a factor of 50 to reach the antiferromagnetic phase not too far below $T_{c}$\cite{G}.  An even larger factor is required to reach the low temperature regime.  In addition, intrinsic heating rate due to spontaneous emission can lead to entropy production as high as 1$k_{B}$ per second\cite{Bloch},  which will destroy many states of interest over the duration of the experiment. It is therefore crucial to invent cooling schemes powerful enough to overcompensate the intrinsic heating. 

The 3D fermion Hubbard model, however, is by no means the most demanding for temperature and entropy reduction.  There is a host of novel cold atom states characterized by very small energy scales. For example, high spin fermions such as  $^{40}K$ have ten Fermi surfaces and can have very remarkable pairing states in optical traps~\cite{HoYip}. Unfortunately, the interactions are so weak that the superfluid transition temperature can be as low as $10^{-11}$K for systems with $10^5$ particles, making the realization of such novel states very difficult. 
Other systems such as large spin Bose gases in optical lattices,  mixtures of quantum gases, and  low dimensional dipolar Fermi gases\cite{Bruun}  
are characterized by similar or even smaller energy scales. Thus, the success of realizing many novel states rests on the ability to reach these unprecedented ultra-low temperature and entropy regimes.

Previously, we proposed a scheme to expel (or ``squeeze" out) the entropy of a Fermi gas in an optical lattice to the surface of the cloud by turning its interior into a band insulator through compression~\cite{HoZhou}. The surface entropy can then be removed by various means.  For example, the surface entropy can be transferred into a surrounding Bose-Einstein condensate (BEC) which can then be evaporated away\cite{HoZhou}.  It is shown that in this way, the entropy per particle can be lowered by a factor of 50.  Other variations of using band insulators to push out the entropy have been considered.~\cite{Bernier,Mat}.  The effective transfer of entropy between quantum gases has also been demonstrated recently~\cite{Ingu}.
It was also pointed out in Ref.~\cite{HoZhou} that this principle of entropy expulsion is not restricted to band insulators, but applicable to all system with an incompressible gapful phase. 

The ``expulsion method"~\cite{HoZhou} and its variations~\cite{Bernier,Mat}, however,  have many limitations.  First of all, many systems of interests  do not have a band insulator or an incompressible phase. This include all quantum gases in single traps {\em without} an optical lattice, as well as  spinor gases and mixtures in optical lattices. 
Secondly, it involves too many steps, requiring first the creation of a band insulator, then the removal of entropy at the surface in a way that will not lead to entropy re-generation, and then the transformation of the band insulator into the state of interest.  Thirdly, its cooling power, though considerable, is still limited. It is not clear whether it can overcome the problem of intrinsic heating.  While in principle one can repeat this process,  the number of steps required makes a repeated running impractical.

In this paper, we present a cooling scheme free of the above limitations.  The scheme is applicable to all quantum gases including mixtures and gases with internal degrees of freedom.  It is designed to extract entropy directly from the region of interest in a continuous way, and hence is capable of combating extrinsic and intrinsic heating in a continuous manner.  This cooling scheme will remain operative as long as the system can equilibrate during the operation. 
However, as temperature drops, the time for equilibration increases and will eventually place a limit on the cooling power of this scheme.  Here, we focus on ways to reach maximum cooling power within adiabatic processes.  How to speed up the equilibration processes will be considered elsewhere. 

\begin{figure}[tbp]
\begin{center}
\includegraphics[width=3in]{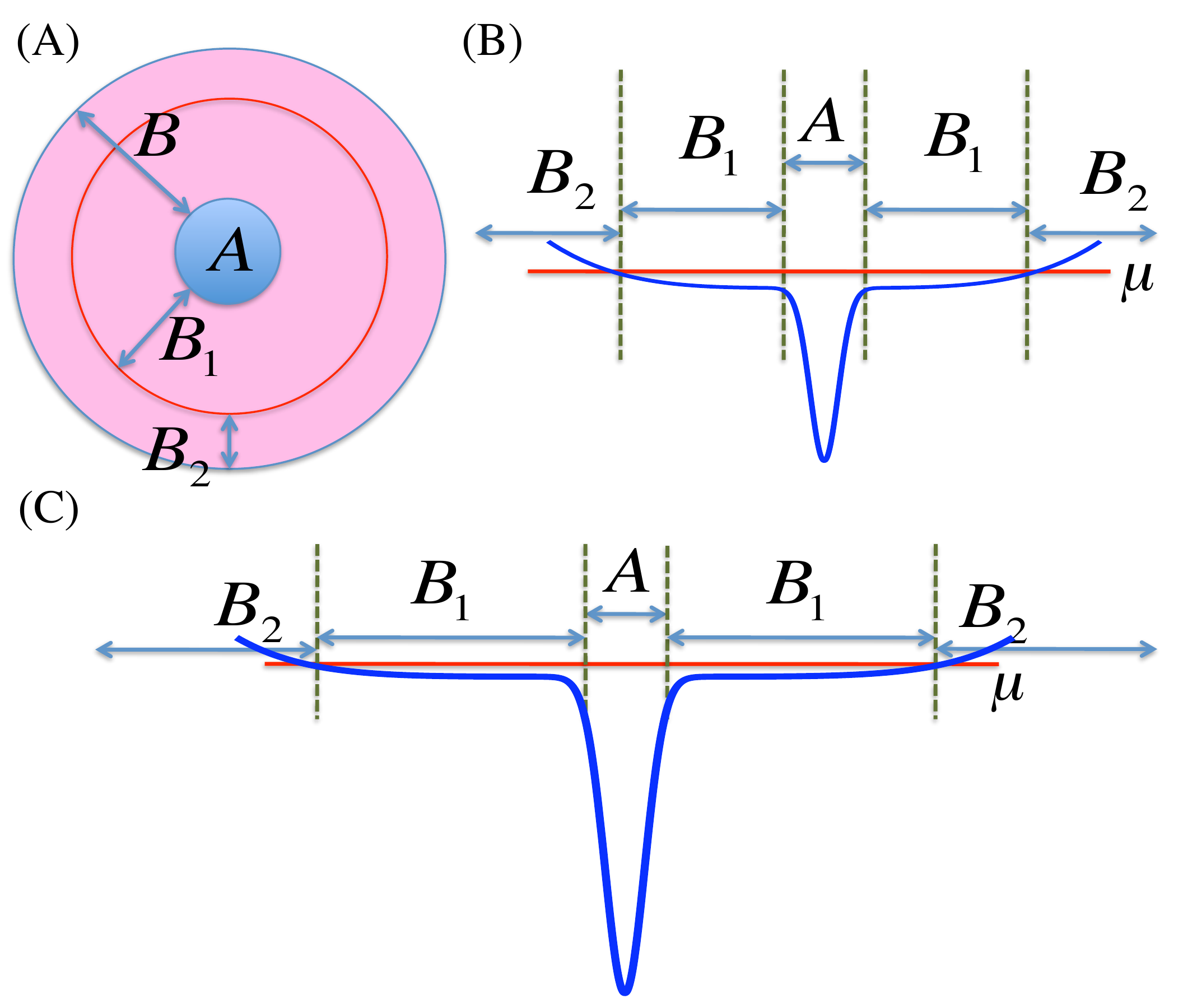}
\end{center}
\caption{Schematic of the entropy extraction scheme: The system of interest is confined in ${\cal A}$. The trapping potential $V({\bf x})$ is represented by the solid blue curve. Chemical potential $\mu$  is represented the red horizontal line. 
As the surrounding region ${\cal B}$ expands, entropy and particles flow out from ${\cal A}$ to ${\cal B}$. 
$\mu$ is kept close to the top of the dimple, and the difference $\tilde{\mu}= \mu-V({\bf 0})$ 
 is maintained roughly constant at the late stage of expansion so that $T/\tilde{\mu}$ decreases as expansion proceeds. 
}\label{fig: fig1}
\end{figure}

{\bf A. Direct entropy extraction:}  The general principle for our scheme is shown in Figures (1A) to (1C). The system of interest is confined in region ${\cal A}$. It is surrounded by an ``extraction region", denoted as ${\cal B}$.  An example of the trapping potential $V({\bf x})$ that gives rise to this situation is shown in Figure (1B). It consists of a box-like potential  (region ${\cal B}$) with an attractive ``dimple" at the center 
(region ${\cal A}$). Region ${\cal B}$ is further divided into regions  ${\cal B}_{1}$ and ${\cal B}_{2}$, 
where $\mu>V({\bf x})$ and  $\mu<V({\bf x})$ respectively.  
If we expand  the volume of ${\cal B}$ adiabatically while keeping that the chemical potential $\mu$ close to the top of the dimple (Fig.(1C)), both energetic particles and entropy will be ``sucked out" from ${\cal A}$ to ${\cal B}$, causing the temperature $T$ to drop.  More importantly, if
 the flow entropy flow out of ${\cal A}$ is faster than the particle flow, then the entropy per particle in ${\cal A}$ (denoted as $(S/N)_{dimple}$) will also decrease. Next, we note that the entropy per particle in ${\cal A}$ is proportional to $T/\tilde{\mu}$, where $\tilde{\mu}= \mu-V({\bf 0})$ is the chemical potential measured from the bottom of the dimple, (see example below). One can therefore reduce $(S/N)_{dimple}$ as ${\cal B}$ expands by keeping $\tilde{\mu}$ at a reasonably high value. 
In addition, we note that 
in the low-fugacity region ${\cal B}_{2}$, the entropy per particle is 
$s/n= k_{B}\left[ 5/2 -  \mu/(k_{B}T)\right]$, 
where $s$ and $n$ are entropy and particle density, respectively. In this region, we have $n=e^{\mu/(k_{B}T)}/\lambda^3 \ll 1$, where $\lambda = h/\sqrt{2\pi M k_{B}T}$ is the thermal wavelength.
Since  in  ${\cal B}_{2}$, $\mu({\bf x})<0$ and $\mu({\bf x})$ increases as $|{\bf x}|$ increases, 
the entropy per particle $s({\bf r})/n({\bf r})$ becomes arbitrarily large as one moves away from the center, making ${\cal B}_{2}$ a highly efficiently region to absorb entropy.
 Our discussion of a box-like potential is for illustration purposes only. In practice, one simply uses a harmonic trap. The  expansion of ${\cal B}$  then corresponds to reducing the trap frequency.

We should mention that the dimple trap has been used before by Ketterle's group to produce a Bose-Einstein condensate (BEC)~\cite{Ketterle}. In that case, the dimple is turned on without changing the harmonic trap.  This actually raises the temperature~\cite{Ketterle}.  Nevertheless, BEC still occurs because $T_{BEC}$ rises even faster due to the rapid increase in phase space density.  This scheme, however, does not lead to large reduction of entropy per particle within the dimple region. To achieve that, it is crucial to open up the harmonic trap while controlling the density at the center of the dimple and the value of the chemical potential as we show below. 

\begin{figure}[tbp]
\begin{center}
\includegraphics[width=2.6in]{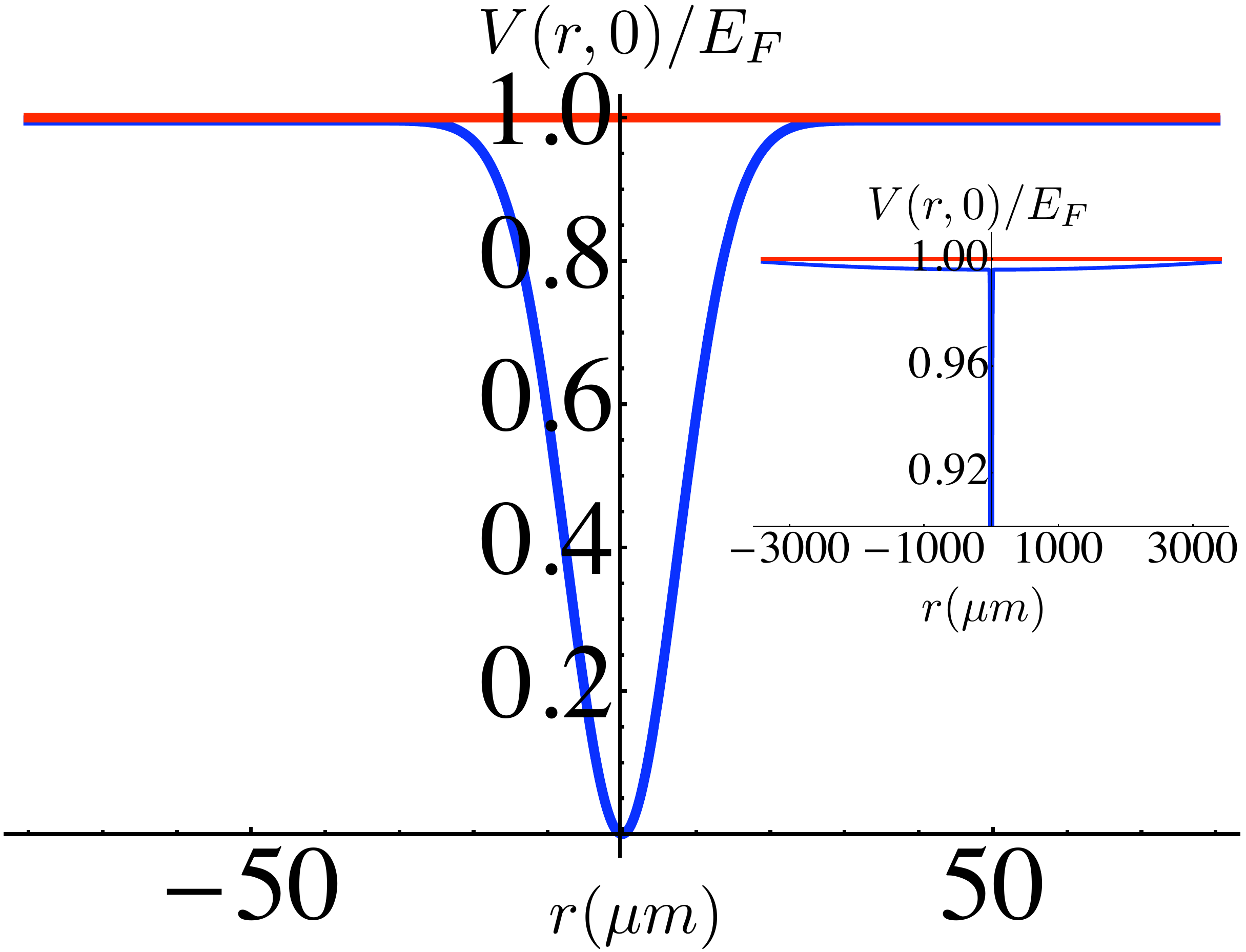}
\end{center}
\caption{Trapping potential for Fermi gas with weak attractive interaction in the $z=0$ plane. The blue curve is the total potential which consists of a dimple and the harmonic potential. The horizontal red line denotes the chemical potential. The inset shows the potential on a larger scale. }\label{fig: fig2}
\end{figure}

{\bf B. Illustration of the Cooling Scheme: } 
To illustrate the cooling power of this scheme, we consider a two-component Fermi gas with weak attractive interactions. At present, superfluidity in Fermi gases can only be achieved in the strongly interacting limit with $|k_{F}a_{s} |\gg 1$, 
where $k_{F}$ is the Fermi wavevector and $a_{s}$ is the $s$-wave scattering length. 
For a weakly attractive gas, $|k_{F}a|\ll1$, the superfluid transition temperature $T_{c}$ is too low to be reached by the cooling methods today.  Here, we show that such ultra-low temperatures are achievable by our simple scheme.  We begin with a (two-component) Fermi gas of $N=1 \times 10^6$ $^6$Li atoms with scattering length $a_{s} = -0.25/k_F$.  The corresponding $T_{c}$ in the bulk is $T_{c}=
0.61E_F e^{-\pi/|2k_{F} a_{s}|}$~\cite{Pethick}, where $E_{F}=\hbar^2 k^2/2M$ is the Fermi energy, 
 (for  $k_{F}= 2.2 \mu m^{-1}$, $T_{c}=0.24$nK. )
Our trapping potential is $V({\bf x}) = V_{h}({\bf x}) + V_{d}({\bf x})$, where $V_{h}$ and $V_{d}$ are the potentials of a harmonic trap and a dimple trap respectively, 
\begin{equation}
V_{h}({\bf x}) =\frac{M}{2}(\omega_{\perp}^2 r^2  +  \omega_{z}^2 z^2), \,\,\,\,\,\,\, 
V_{d}({\bf x}) =  V_{o} ( 1-e^{-r^2/D^2}).  
\label{V} \end{equation}
Here,  ${\bf r}=(x,y)$, ${\bf x}=(x,y,z)$;  $\omega_{\perp}$ and $\omega_{z}$ are the frequencies of the harmonic trap; $V_{o}$ and $D$ are the depth and the width of the dimple.  We have chosen a dimple with no $z$-variations to mimic a red de-tuned laser along $z$. (The specific form of the dimple, however, is not important for our discussions). We have added a constant to the dimple potential so that the minimum value of the total potential is zero.

We consider an initial state where the Fermi gas is confined in a harmonic trap without the dimple, 
 \begin{equation}
V^{(i)}({\bf r}) =\frac{1}{2}M\omega^{(i) 2} (r^2  +  z^2), \,\,\,\,\,\,\,  V_{d} =0; 
\end{equation}
with entropy per particle $(S/N)^{(i)}= 0.45$, which is close to the lowest value achievable in a single trap today.  
For $\nu^{(i)}=\omega^{(i)}/ 2\pi=23.7$Hz, this entropy corresponds to an initial temperature $T_{i}= 9.4 nK$, and a Fermi momentum at the center of the trap $k_{F}=2.2 \mu m^{-1}$. 
Next, we turn on the dimple and open up the harmonic trap adiabatically. The depth and width of the dimple are adjusted so that density (and hence $k_{F}$) at the center of the dimple remain constant during the entire process. 
The final configuration of the trapping potential $V^{(f)}$ is shown in Figure 2 at two different length scales. 
The potential depth $V_{o}$ is measured in units of the Fermi energy $E_{F}$ at the center of the dimple, which is fixed to its initial value $E_{F}=\hbar^2k_{F}^2/2M$.  The frequency of the dimple trap near its bottom is $361$Hz. (See also Table 1.) The size of the dimple is $D=10.6\mu m$.   The horizontal straight line indicates the value of the chemical potential $\mu$.

 

\begin{figure}[tbp]
\begin{center}
\includegraphics[width=3in]{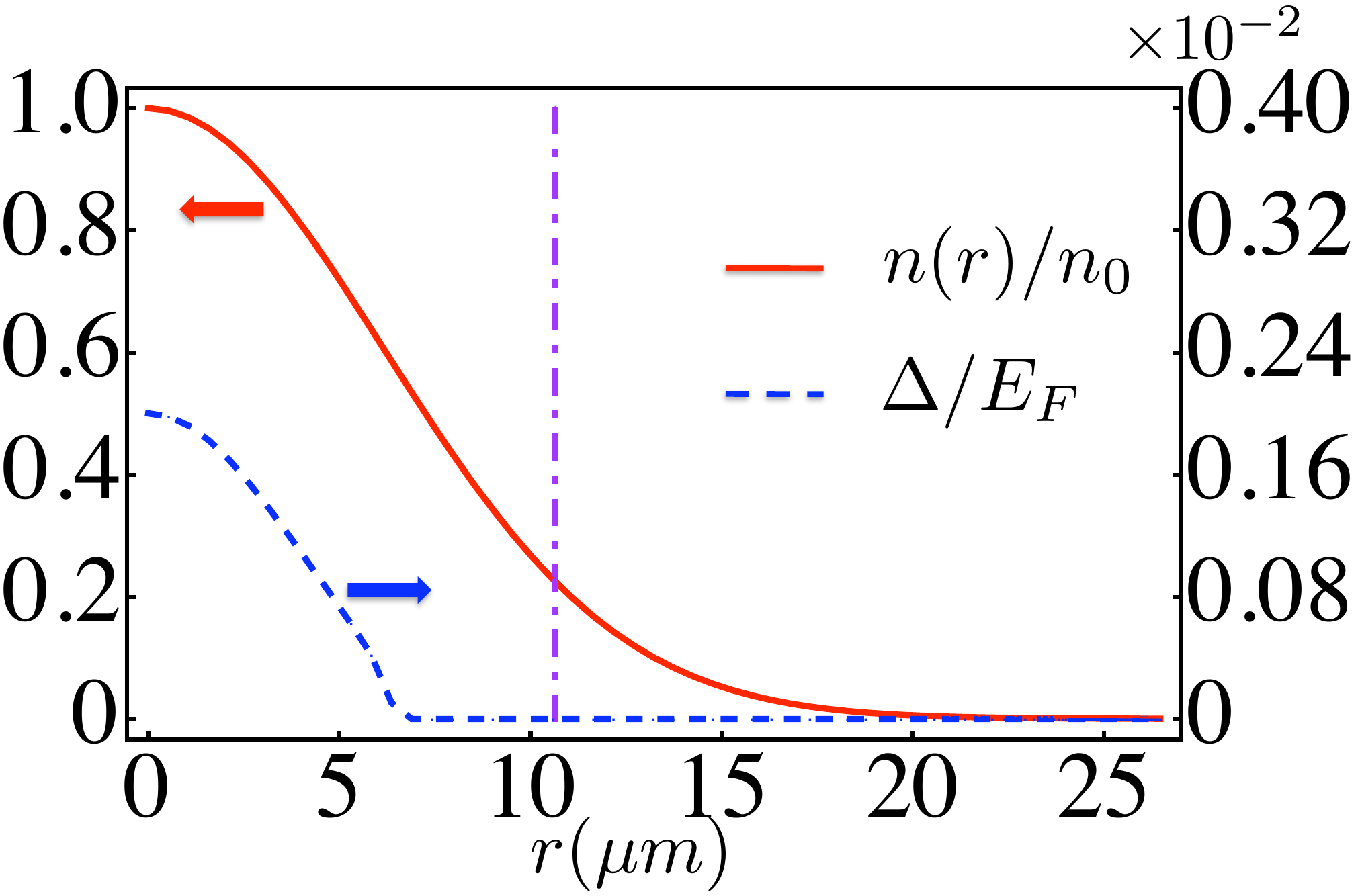}
\end{center}
\caption{Density and energy gap of the Fermi gas in the potential shown in Figure 2.  The parameters of this system are given in Table 1.  }\label{fig: fig3}
\end{figure}

Using weak-coupling BCS theory, one can obtain the number density $n$, entropy density $s$, and energy gap $\Delta$  as functions of $\mu$ and $T$. (See {\bf Appendix}.)  We can then calculate these quantities and the temperature within the local density approximation as $V({\bf r})$ is varied.  The properties of the system  in the final potential shown in Figure 2 are given by Figures 3 to 5, as well as in Table I.  

\begin{tabular}{||c|c||}
\hline
$\nu^{(i)}_{\perp}=\nu^{(i)}_{z} = 23.7 Hz $  & 
$\nu^{(f)}_{\perp}= 0.1 Hz$, \,\,\,\, $\nu^{(i)}_{z} = 1 Hz $  \\
\hline
$V_{o}^{(i)}=0$  & $V_{o}^{(f)}=0.996 E_F$\\
\hline
$\mu^{(i)} =206  nK$  & $\mu^{(f)}= 206 nK$ \\
\hline
$k_{F}^{(i)} =k_{F}^{(f)} =2.2 \mu m^{-1}$  & $k_{F}^{(i)}a_{s}=k_{F}^{(f)}a_{s}  =-0.25 $  \\
\hline
$T^{(i)} =9.4 nK$  & $T^{(f)}= 0.047 nK= T_{c}/5$ \\
 \hline 
$(S/N)^{(i)}= 0.45 $  & $(S/N)^{(f)}_{dim, SF} = 0.0005$ \\
\hline
  &   $D=10.6 \mu m$ \\
 \hline
\end{tabular}

\vspace{0.2in}

\begin{figure}[bp]
\begin{center}
\includegraphics[width=3in]{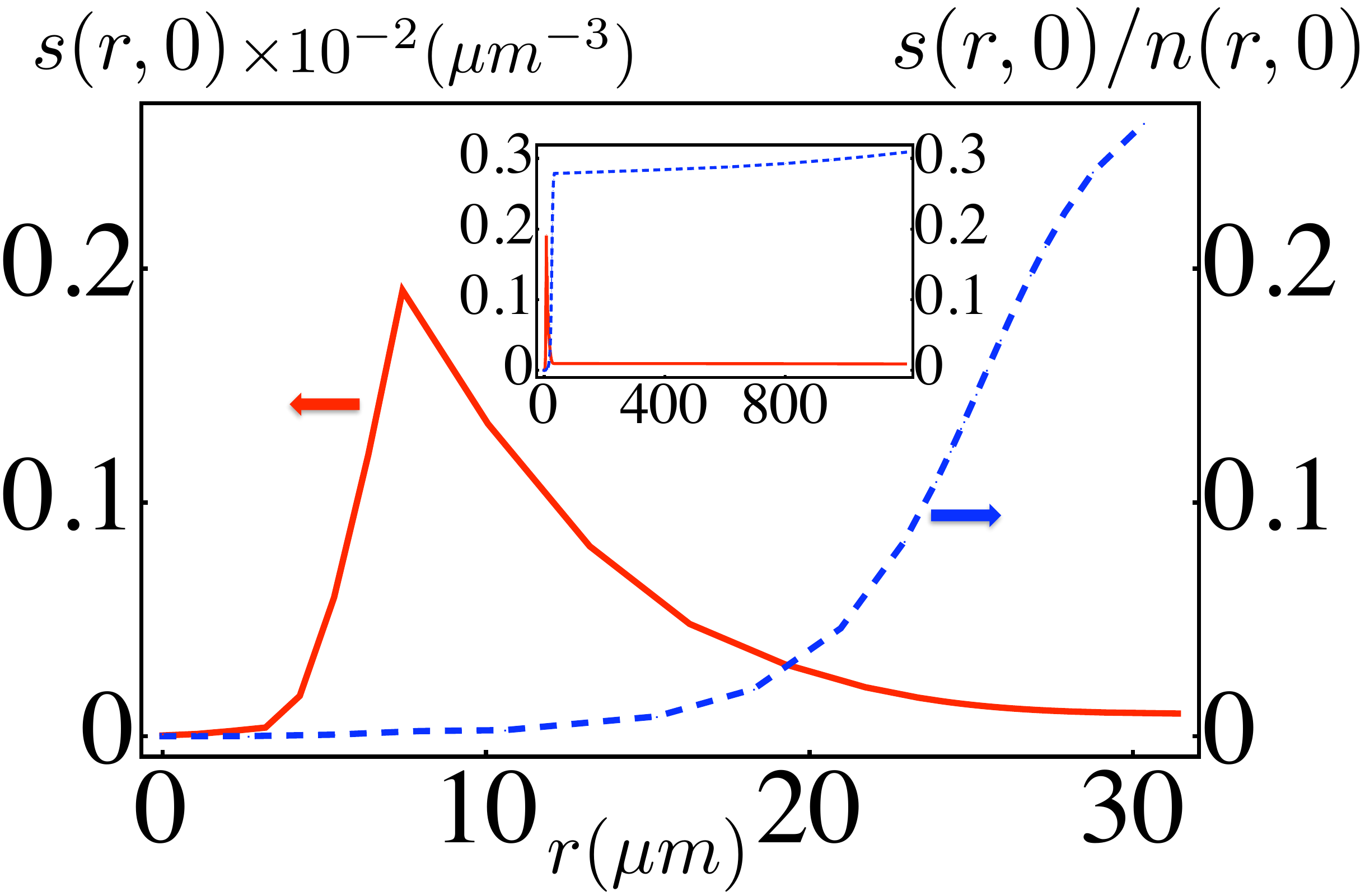}
\end{center}
\caption{Entropy density $s(r)$ and local entropy per particle $s(r)/n(r)$ for the Fermi gas shown in Figure 3 are represented by the red and blue curves. The inset is the same figure in much larger scale, with the same labeling of the axes. 
 }\label{fig: fig4}
\end{figure}

Figure 3 shows the density profile and the gap. For these trap parameters, the temperature has been reduced to one-fifth $T_{c}$, and pair condensation has taken place inside the dimple.  Figure 4 shows the entropy density $s({\bf r})$ and entropy per particle $s({\bf r})/n({\bf r})$ as a function of $r$ in the $z=0$ plane at two different scales.  
The high entropy density accumulated at the edge of the dimple is due to the rapid drop of density there. 
The entropy per particle in the dimple region remains very low.  It is more illuminating to look at the total particle number $N(R)$ and total entropy $S(R)$ contained within a region of radius $R$, 
$S(R)= \int_{0}^{R}  (2\pi r {\rm d}r ) \int {\rm d}z s(r, z)$, 
$N(R)= \int_{0}^{R}  (2\pi r {\rm d}r ) \int {\rm d}z n(r, z)$; 
which are plotted in Figure 5.  We find that the dimple $(R<D)$ contains about 0.15 of total number of particles, whereas essentially all the entropy resides outside the dimple.  Counting the total entropy and particle number in the dimple region (consisting of both normal and superfluid components), we find that their ratio is $(S/N)^{(f)}_{dim} = 0.0014$, reduced from the original value by a factor of $320$.  Counting only the superfluid region within the dimple, the ratio $(S/N)^{(f)}_{dim, SF} $ is 900 lower than the initial value, (see Table  1).  We stress that this reduction is by no means the limit.  More cooling can be obtained by further opening the harmonic trap and adjusting the dimple potential. 

\begin{figure}[tbp]
\begin{center}
\includegraphics[width=3in]{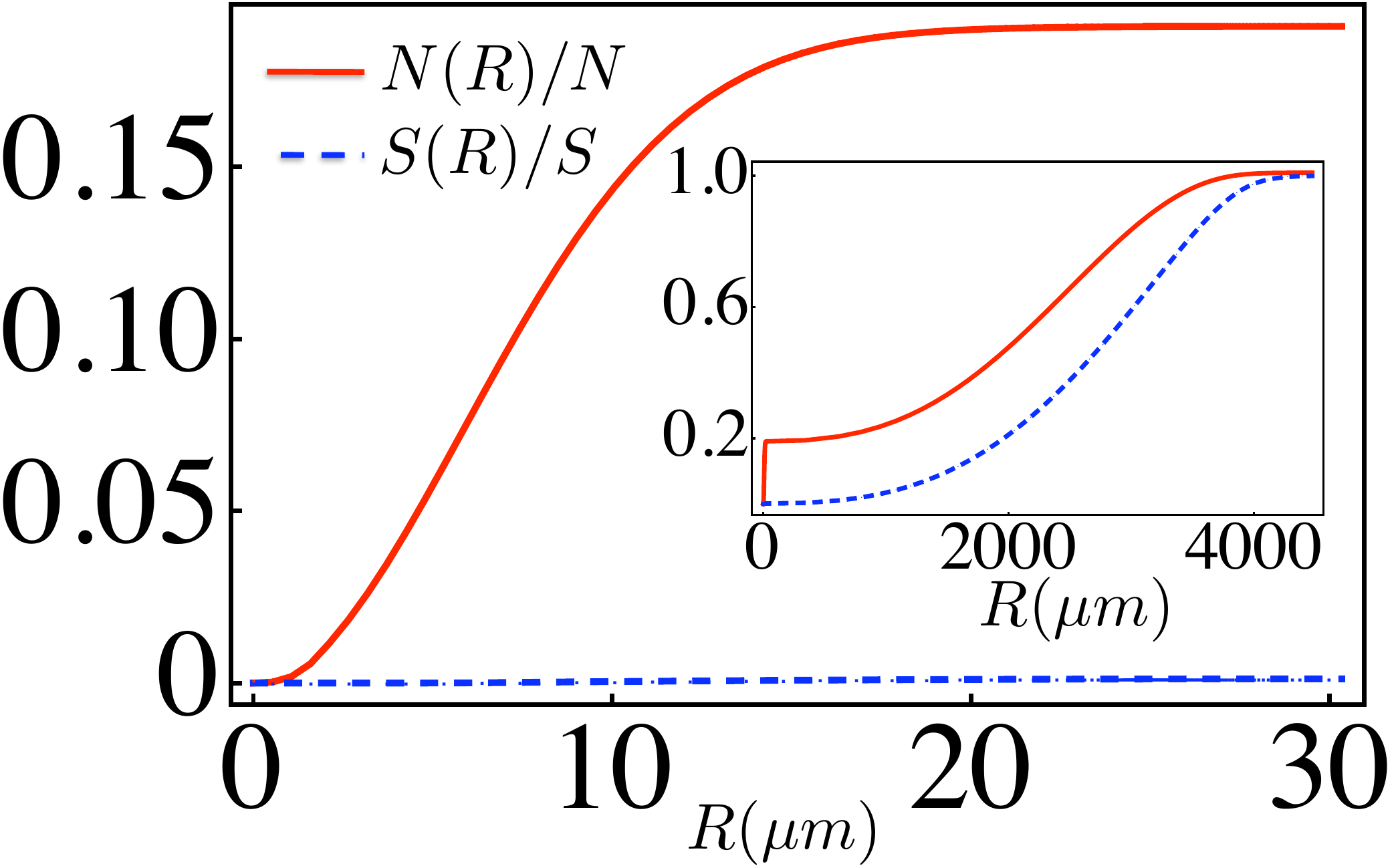}
\end{center}
\caption{Total particle number and total entropy within a distance $R$ from the center. }\label{fig: fig5}
\end{figure}

{\bf C. Equivalence to   continuous evaporation: }  As the potential outside the dimple becomes flatter and flatter,  or more generally,  as the volume of the region ${\cal B}$ in Figure 1 continues to grow,  the system inside the dimple (region ${\cal A}$ in Figure 1 ) can not tell whether the potential outside is  expanding but eventually bounded, or a flat potential that turns downward at large distances, 
 allowing particles to leak out to infinity. In other words, as the trap  opens up adiabatically, as far as the system in ${\cal A}$ is concerned, it amounts to a process of continuous evaporation.  From this viewpoint, one can see another advantage of this scheme: the continuous removal of entropy in ${\cal A}$ allows one to combat extrinsic or intrinsic  heating.  By continuous adjusting the depth $V_{o}$ and width $D$ of the dimple, one can keep the chemical potential close to the top of the dimple, therefore continuously ``evaporating" away the hot particles inside. 

Our scheme differs, however, from the usual evaporative cooling in a significant way. 
In the evaporation scheme used by current experiments, one lowers the potential near the edge of the cloud to let the hottest particles escape~\cite{Jin}.  While temperature is lowered in this process,  the density at the center of the trap (and hence $E_{F}$) is also decreased.  As a result, once $T$ falls below $E_{F}$,  there is no significant reduction in the ratio of  $T/E_{F}$~\cite{Jin}, and hence no significant reduction on entropy per particle in the dimple, since $(S/N)\sim T/E_{F}$ for a Fermi gas.   In our case, the density at the center of the dimple is {\em kept fixed} during the evaporation process. The ratio $T/E_{F}$ drops rapidly as the cooling process proceeds. This simple step makes an enormous difference in the cooling power of the evaporation process.  

Our discussions assume that the process is adiabatic. As mentioned in the opening, as temperature drops, it will take the system in the dimple longer and longer to equilibrate.  Eventually, it will reach a point where adiabaticity can not be maintained.  The equilibration time, which is system-dependent, is the intrinsic limit of the cooling power of this scheme.  The speeding up of relaxation processes will be studied elsewhere. 

Finally, we point out that the principle discussed in {\bf A}  applies to all quantum gas systems.  For lattice quantum gases, Eq.~({\ref{V}) will be a potential in addition to the lattice potential.  
The crucial steps remains the same: relaxing the harmonic trap adiabatically, adjusting the dimple so that $\mu$ remains close to the top, and $\tilde{\mu}$ is kept at a value necessary to realize the state of interest. Despite the simplicity of these steps, they make big differences from previous studies\cite{Ketterle,Jin}.  The physics contained in these steps can work magic, enabling entropy per particle to be lowered from the current values by orders of magnitude.


{\bf Appendix:}  Within BCS theory, number density $n$ and entropy density $s$ are given by
$n=$ $\Omega^{-1} \sum_{\bf k}$ $\left(1- \frac{ \xi_{\bf k} }{E_{\bf k}}\right)$, 
$\xi_{\bf k}$ $=$ $\epsilon_{\bf k}-\mu$, 
$E_{\bf k}$$=$$\sqrt{ \xi_{\bf k}^{2}+\Delta^{2}}$,
$\epsilon_{\bf k} = \hbar^{2}k^{2}/(2M)$, 
$s$$=$$ \Omega^{-1} \sum_{\bf k}$$\left[  (1-f(E_{\bf k})){\rm ln}(1-f(E_{\bf k})) + f(E_{\bf k}){\rm ln}f(E_{\bf k})\right]$,
where  $\Omega$ is the volume of the system, $f(x)= 1/(e^{x}+1)$  is the Fermi function, and the gap $\Delta$ is determined by 
\begin{equation}
\frac{M}{4\hbar^2\pi{a}}=-\frac{1}{2}\sum \left(\frac{\tanh {E}_{\bf k}/(2k_{B}T)}{E_{\bf k}}-\frac{1}{\epsilon_{\bf k}}\right).
\end{equation}
Within the local density approximation, the total particle number $N$ and total entropy $S$ are given by $N=\int n(\mu - V({\bf x}, T))$ and  $S=\int s(\mu - V({\bf x}, T))$.  For given values of $N$ and $S$, one can then find $\mu$ and $T$, which then gives the results in Figs.~3-5. 

We would like to thank Tilman Essingler, Leticia Tarruell, Randy Hulet, and Jun Ye for helpful discussions.  This work is supported by NSF Grant DMR-0907366, and DARPA under the ARO Grant No. W911NF-07-1-0464.

\end{document}